**Electronic health record phenotyping improves detection and screening of type 2 diabetes in the general United States population: A cross-sectional, unselected, retrospective study**

**Ariana E. Anderson[1], Wesley T. Kerr*[1,2], April Thames[1], Tong Li[1], Jiayang Xiao[1], Mark S. Cohen[1]**
  (1) Department of Psychiatry and Biobehavioral Sciences, University of California, Los Angeles
  (2) Department of Biomathematics, David Geffen School of Medicine at UCLA
   (*) Corresponding author
Contact Information:
Semel Institute
760 Westwood Plaza, Ste B8-169,
Los Angeles, CA 90095-1406
Phone: (310) 254-5680
Email: WesleyTK@UCLA.edu

**Abstract:**

Objectives: In the United States, 25% of people with type 2 diabetes are undiagnosed. Conventional screening models use limited demographic information to assess risk. We evaluated whether electronic health record (EHR) phenotyping could improve diabetes screening, even when records are incomplete and data are not recorded systematically across patients and practice locations.
Methods: In this cross-sectional, retrospective study, data from 9,948 US patients between 2009 and 2012 were used to develop a pre-screening tool to predict current type 2 diabetes, using multivariate logistic regression. We compared (1) a full EHR model containing prescribed medications, diagnoses, and traditional predictive information, (2) a restricted EHR model where medication information was removed, and (3) a conventional model containing only traditional predictive information (BMI, age, gender, hypertensive and smoking status). We additionally used a random-forests classification model to judge whether including additional EHR information could increase the ability to detect patients with Type 2 diabetes on new patient samples.
Results: Using a patient's full or restricted EHR to detect diabetes was superior to using basic covariates alone (p<0.001). The random forests model replicated on out-of-bag data. Migraines and cardiac dysrhythmias were negatively associated with type 2 diabetes, while acute bronchitis and herpes zoster were positively associated, among other factors.
Conclusions: EHR phenotyping resulted in markedly superior detection of type 2 diabetes in a general US population, could increase the efficiency and accuracy of disease screening, and are capable of picking up signals in real-world records.

## Introduction

Although roughly 25% of those with type 2 diabetes are undiagnosed in the United States, population-wide screening for diabetes currently is not cost-effective because of the additional time and laboratory testing required[1-3]. Intervention studies have shown that diabetes can be prevented in high-risk individuals[3-5], while weight loss and lifestyle changes can revert the recently diagnosed patients (<4 years) to pre-diabetes status[6]. This makes population-wide screening not just an issue of prevention, but also one of treatment.

The total estimated cost of diagnosed diabetes in 2012 reached a staggering $245 billion, a 41% increase since 2007. People with diagnosed diabetes, on average, have medical expenditures approximately 2.3 times higher than people who do not have diabetes[7]. Given that expenditures for patients with diabetes account for 20% of healthcare spending annually,[7] and that new legislation bars insurers from dropping coverage of ill patients, characterizing diabetes risk using electronic health records (EHR) allows this financial liability to be both appreciated and mitigated by those bearing the financial burden of the disease.

Diabetes screening risk scores combine both patient demographic information and laboratory testing to predict future likelihood of developing diabetes. Risk scores are derived from demographic information such as age, gender, comorbid hypertension, ethnicity, and body mass index (BMI). Laboratory tests include fasting plasma glucose concentration, oral glucose tolerance test or hemoglobin A1c (compared more thoroughly in [8]). These tests often require fasting, routine blood draws, and monitoring, which can place undue burden on the patient, staff, and treating physician. This makes population-wide screening expensive and impractical, particularly in resource limited health-care settings which are the most likely to service at-risk patients[2, 3]. Laboratory tests cannot, and should not, be performed on every individual after every health care encounter.

Diabetes screening is recommended by the U.S. Preventive Services Task Force only for asymptomatic adults with treated or untreated blood pressure over 135/80 mm Hg, even though hypertension is only one of many known risk factors for diabetes[9]. In this sample, 27.5% of patients with type 2 diabetes did not have a hypertension diagnosis; this criterion would result in lab tests for 1 in 3 healthy patients, while failing to identify roughly 1 in 4 patients with type 2 diabetes. These data suggest that more sophisticated screening methods are needed, consistent with the Wilson and Jungner criteria[10, 11].

Many current diabetes risk models are not generalizable to the disproportionate number of ethnic/racial minorities with Type 2 diabetes in the U.S.[12]. For example, the FINDRISC score was developed on Finnish patients[13]. Although it was found to be the best risk assessment tool in Caucasian patients[8], it was suboptimal for use on Arab and Filipino populations[14, 15]. Therefore, wide-scale screening methods that assume a "one-size-fits all" approach is simply not feasible among heterogeneous populations. Considering that African Americans, Hispanic/Latino Americans, American Indians, Asian Americans, and Pacific Islander Americans are at particularly high risk for type 2 diabetes[1], it is important to understand the unique risk factors that continue to drive this growing health disparity among these groups, and to assess risk accurately in the subpopulations most likely to have the disease.

Electronic health care records are used by more than half of the nation's health care providers, and more than 80% of hospitals have implemented electronic records since the 2009 stimulus bill[16]. EHRs are also of interest among research investigators, as this provides a rich data resource that can improve research efficiency, as advocated by the NIH's Health Care Systems

Collaboratory (https://www.nihcollaboratory.org/) program which engages healthcare systems to improve medical research[17]. EHR-based phenotypes can profile individuals who may benefit from interventions[18]. In a similar fashion, an evaluation guided by EHRs may improve the treatment and prognosis for patients with type 2 diabetes[19]. For instance, usage of an EHR was associated with a decreased rate of emergency department visits in individuals with diabetes[20, 21], and EHR data have been used to compute the prospective risk of developing dementia in individuals with diabetes[22]. While EHRs have demonstrated potential for detecting and monitoring diabetes[23-26], previous studies have used only a subset of all information available in the medical record, and typically have assessed risk only on patients for whom there were specific laboratory results available (*e.g.*, fasting plasma glucose). In order for EHR phenotyping to be used in practice, it must improve accuracy even when using "typical" quality records, containing large amounts of missing data and unsystematic recordings across practice locations, consistent with at-risk patients receiving inconsistent medical attention[27].

Current risk scores pose an economic and logistical challenge for population-wide screening, and are less sensitive to detecting diabetes in non-white populations who are more likely to have diabetes[16,17]. More generally, automatically predicting chronic disease using electronic health records has numerous applications outside of clinical practice, and would open the risk-assessment doorway to those who ultimately bear the financial burden incurred by the disease: the insurers. A better estimate of risk on the part of insurers could encourage targeted patient-education and incentive programs to reduce financial liability.

EHR models extend screening, conventionally framed between the doctor and the patient, to the payer and the patient. Data mining methods are powerful, but wild-type electronic records frequently are messy[18]; these tools should be validated against real-world data if realistic results are desired. We examine whether augmenting risk scores using EHR-derived phenotypes would increase sensitivity in the general population for detecting patients who should be screened further using laboratory testing, even when records are incomplete, and are not recorded systematically across health professionals and/or practice locations. When implemented on a population, this step-wise screening process would decrease the public health cost of more expensive testing, while simultaneously identifying previously overlooked at-risk patients.

**Subjects**

The study population included 9,948 patients, 18.1% diagnosed with type 2 diabetes, within the EHRs provided by the web-based company, Practice Fusion. Practice Fusion provides an EMR service which is free to health care providers. These records were collected between 2009 and 2012 in a retrospective study, from 1,137 unique sites spanning all 50 United States, detailed further in Table 1. This dataset is public and de-identified. We intentionally use an unselected patient population who had a wide variety of laboratory tests, prescribed medications, and diagnoses across 1,137 different practice locations. In contrast to a controlled research environment[28], this EHR data approximates the diverse US clinical environment, including the presence of unsystematic and incomplete information reflective of at-risk patients receiving inconsistent medical care.

Exact visit dates were removed as patient-identifying information. A patient was labeled as either healthy, or having "Type 1" or "Type 2" diabetes according to at least one corresponding diagnosis within ICD-9 250.X category (no patients had mixed diagnoses). No ethnicity information was provided in this sample. There were approximately 131,000 patient unique transcript entries (containing height, weight, BMI, BP, Respiratory Rate, and Heart Rate) for the

9,948 patients across an approximately 4 year time period, leading to roughly 2-3 visits/year/patient, on average.

**Table 1: Demographic Information**

|  | Healthy Control | Type 1 Diabetes | Type 2 Diabetes |
|---|---|---|---|
| n | 7978 | 165 | 1805 |
| Male (%) | 40.60% | 51.50% | 50.60% |
| Age | 50.81 (17.65) | 55.67 (15.40) | 62.88 (13.04) |
| BMI | 28.89 (6.07) | 29.00 (6.52) | 29.15 (6.02) |
| Systolic BP (mm Hg) | 126.24 (17.76) | 127.74 (19.35) | 127.13 (18.57) |
| Diastolic BP (mm Hg) | 76.75 (11.00) | 77.14 (12.22) | 77.04 (11.14) |
| Total Medications Prescribed | 4.53 (4.52) | 3.98 (3.98) | 4.25 (4.58) |
| Total Risk Factors | 0.71 (0.91) | 1.06 (.89) | 1.23 (.93) |
| Hypertension DX (%) | 34.50% | 64.20% | 72.50% |
| High Cholesterol (%) | 28.70% | 51.50% | 62.40% |
| Smoking (%) | 6.30% | 5.40% | 5.40% |

This dataset was rich in the breadth of information it contained: lab results; medication dosage, and history; basic patient demographic information (age, gender, state); smoking status; transcripts (BMI, systolic/diastolic blood pressure (BP), height, weight); allergies; and diagnosed conditions for each visit as ICD-9 codes. We identified and condensed redundant features manually (*e.g.* Warfarin and Coumadin). When BMI values were over 70, or below 10, we re-coded height and weight as not available ("-NA-").  Moreover, the "Healthy Controls" in this sample, on average, had more prescribed medications and higher smoking rates than patients with Type 1 or Type 2 Diabetes.  This natural sample provided a more realistic control group to identify which factors in the EHR were predictive of a Type 2 diabetes diagnosis.

Missing data were common in this sample: less than 1% of the patients had a recorded family history of diabetes (ICD9 V18.0), despite a prevalence of 11.8% in the US population. This posed a "worst case" scenario for prediction: given missing, unsystematic and incomplete information from a patient's medical history, could residual information still augment current diabetes risk scores in a way that improves the accuracy and efficiency of type 2 diabetes screening in the general population?

**Materials and Methods**

We assessed whether Type 2 diabetes risk scores could be improved with EHR phenotypes, created using the additional medical and diagnostic information contained in the EHR. Because the visit dates were removed to protect patient privacy, longitudinal data were not available.  It is therefore unknown whether patients developed diabetes during their time of service, or whether it preceded their entry into this study.  It is similarly unknown whether patients identified as "healthy" had undiagnosed diabetes.  Similarly, the ordering of medications, non-diabetes diagnoses, and the diabetes diagnosis are similarly unknown.  Using real-world clinical data, these models then assess the current likelihood of a patient having a current diagnosis of Type 2 Diabetes, rather than the future likelihood of developing diabetes. The value of including EHR information was computed by comparing models using a chi-square test.

We predicted current Type 2 diabetes status using a multivariate logistic regression in R[29] comparing three separate models: (1) Conventional Model mimicking conventional risk scores; (2) a full "EHR Model" based upon the EHR phenotype, containing conventional information and both diagnostic and prescription information; (3) "EHR DX" model which contained conventional information and diagnostic information. Within this model, prescription information was removed because a diabetes diagnosis would likely physician prescribing habits. Models are described in more detail below.

The first model (Conventional model) mimics conventional risk scores by including only the limited subset of covariates (smoking status, gender, age, BMI, and hypertensive status) that have been used in current diabetes risk models [8, 21], and included all interaction effects. This was used as a reference standard in lieu of established risk scores because it ensures that difference between the two models is attributable to the structure and covariates, instead of the underlying study populations. We compared these models in a hierarchical regression with a chi-square test, and computed Receiver Operating Characteristic (ROC) curves by measuring the Area Under the Curve (AUC) within the R package ROCR [30].

The second "EHR Model" used 298 features: 150 most common diagnoses, 150 most commonly prescribed medications (before condensing name-brand and generic), transcript information (Table 2), and other specialized features summarized in Table 2. Hypertensive status, hyperlipidemia, and metabolic diagnosis all were assessed separately. In order to reduce bias we removed as predictors established treatments and complications of diabetes mellitus: primary and secondary diabetes-related diagnoses (ICD-9 250.X, 249.X), foot ulcers (ICD-9 707.X), diabetic retinopathy (ICD-9 362.01), polyneuropathy in diabetes (ICD-9 357.2), diabetic cataract (ICD-9 366.41), and diabetes mellitus complicating pregnancy, childbirth or the puerperium (ICD-9 648.0X). Medications used to treat diabetes, such as metformin (Glucophage™), were excluded from the model.

For the EHR Model, we created an "additional risk factor" variable tallying common comorbidities of diabetes, including conditions that have been shown to be more common in diabetes, but may be caused by factors other than diabetes. These included: candidiasis of skin and nails, malignant neoplasm of pancreas, other disorders of pancreatic internal secretion, polycystic ovaries, disorders of lipoid metabolism, overweight, obesity, and other hyperalimentation, trigeminal nerve disorders, hypertensive heart disease, acute myocardial infarction, other acute and subacute forms of ischemic heart disease, old myocardial infarction, angina pectoris, other forms of chronic ischemic heart disease, atherosclerosis, gingival and periodontal diseases, disorders of menstruation and other abnormal bleeding from female genital tract, unspecified local infection of skin and subcutaneous tissue, acquired acanthosis nigricans and tachypnea.

The third "EHR DX Model" used all features as in the EHR model, yet excluded all the medications. This was based on the assumption that a clinician's prescribing behavior would likely be influenced by a patient's diabetes status. Given that this sample is not longitudinal, excluding medication information reduces the bias inherent in all observational studies.

Table 2: Summary of Variables Used to Predict Diabetes

| Variable | Description |
|---|---|
| Demographic Information | age, gender |
| Transcript information | Systolic/Diastolic BP, Height, Weight, BMI |
| Diagnosis | 150 most common ICD9 Codes (primary digits, i.e. 152.X ) |
| Medications | 150 most commonly prescribed |
| Hypertension Diagnosis | ICD9 codes 401-405 |
| Cholesterol Diagnosis | ICD9 code beginning with 272 |
| Metabolic Diagnosis | ICD9 Code 277.7 |
| Diagnosis | Total number |
| Acute Diagnosis | Total number |
| Additional Risk Factors | Diagnoses related to diabetes |
| Family Metabolic Disorder History | ICD9 code of V18.1, V18.11, V18.19 |
| Abnormal Lab Results | Total number |
| Percent Abnormal Lab Results | |
| Geographical Region | Defined by http://www.census.gov/geo/maps-data/maps/pdfs/reference/us_regdiv.pdf |
| State Diabetes Rate | 2010 National Diabetes Surveillance System Location Diabetes Rate |
| Family Diabetes Risk | ICD9 V18.0 |
| Family Metabolic Risk | ICD9 V18.1, V18.11, V18.19 |
| Family Diabetes History | ICD9 code of V18.0 |
| High Risk Practice | Average diabetes rate is above 8.3% at that practice location |
| Smoking Status (NIST Codes) | Current, Former, Never Smoked, NA |

Within the EHR model, the multivariate logistic regression implies necessarily that the covariates of interest and the resulting *p*-values are statistically dependent, making traditional multiple hypotheses correction methods unnecessary (*e.g.* Bonferroni correction). We make our interpretation more conservative by adjusting for the false discovery rate (FDR) using the graphically sharpened method, setting the maximum proportion of false discoveries at .05[31, 32] as implemented by Pike.[33] The FDR is computed over all 298 estimated *p*-values. We indicate which variables had *q*-values (adjusted *p*-values) in the FDR significance range using an asterisk within the regression tables, along with odds-ratio 95%-confidence intervals which take into account the prevalence of the risk factors being considered.

Finally, we assessed the validity of these three models with a hold-out dataset using a random forests prediction model, to assess the sensitivity of these models to overfitting [34]. In the random forests model, decision trees are constructed by resampling with replacement from the data and the predictors. Prediction of cases, using individual decision trees which did not sample that observation, provides an unbiased estimate of the testing accuracy. We run the random forest model using the data from the EHR model, the EHR DX model, and the conventional model. Machine learning methods such as random forests are undeniably more powerful in their prediction, and the testing accuracy they provide is an unbiased estimate of the predictive power on new observations. Here, we create random forests models using a regression, to create similar ROC curves as the logistic models.

## Results

The Full EHR model had sensitivity: 80.6% (78.6%-82.4%), specificity: 74.0% (72.9%-74.9%), and overall accuracy: 75.2% (77.3%-79.0%). When excluding medications in the EHR model, the overall accuracy dropped only slightly to 73.2% (72.3%-74.1%). The DX EHR model had sensitivity: 79.2% (77.2%-81.0%), specificity: 71.9% (70.9%-72.8%). The conventional model had sensitivity: 78.1% (76.1%-80.0%), specificity: 60.2% (59.1%-61.3%), and overall accuracy:

63.4% (62.4%-64.4%). All ranges correspond to 95% confidence intervals, using a decision threshold corresponding to the disease prevalence in the population (18.1%). All ranges reflect 95% confidence intervals.

The Full EHR model had a positive predictive value of 40.7% (39.0%-42.3%), and a negative predictive value of 94.5% (93.9%-95.1%). The DX EHR model had a positive predictive value of 38.4% (36.9% - 40.0%) and a negative predictive value of 94.0% (93.3%-94.5%). The conventional model had a positive predictive value of 30.3% (28.9%-31.7%) and a negative predictive value of 92.5% (91.8%-93.3%).

The Full EHR model and the DX EHR Models both predicted better than the conventional models (chi-square test, p<0.001). For the Full EHR Model, the AUC was 84.9%; for the DX EHR Model, the AUC was 83.2%; for the Conventional model, the AUC was 75.0%. The sensitivity of the EHR models was uniformly superior for all thresholds, as shown in Figure 1. Factors associated significantly with a Type 2 diabetes diagnosis for the EHR model are presented in Table 3, with the full EHR model containing 298 coefficients and significance levels supplied in the Supplemental Table along with the Conventional Model.

**Table 3: Statistically Significant Covariates to determine Type 2 Diabetes Status using EHR phenotypes:** variables on the top half of the table decrease the likelihood of type 2 diabetes, while variables on the bottom half increase the likelihood of type 2 diabetes. Significance indicates * for P < 0.05, ** for P < 0.01, and *** for P < 0.001. Q-significance indicates p-values after adjusting for the false-discovery rate, which controls for the total rate of falsely rejected null hypotheses.

| Summary | Odds Ratio | CI 2.5% | CI 97.5% | Pr(>|z|) | P-Significance | Q-Significance (FDR) |
|---|---|---|---|---|---|---|
| (Intercept) | 0.00 | 9.71E-04 | 1.50E-02 | 0.00000 | *** | * |
| ICDV77: Special screening for endocrine nutritional metabolic and immunity disorders | 0.13 | 3.45E-02 | 3.45E-01 | 0.00030 | ** | * |
| Depot medroxyprogesterone: Contraceptives, Hormones/antineoplastics, Progestins | 0.20 | 6.25E-02 | 5.03E-01 | 0.00191 | ** | * |
| Diazepam: Benzodiazepine | 0.26 | 8.00E-02 | 6.99E-01 | 0.01449 | * | |
| Klonopin: Benzodiazepine | 0.31 | 8.44E-02 | 8.57E-01 | 0.04129 | * | |
| Alaska, Hawaii, California, Oregon, Washington Resident | 0.42 | 2.29E-01 | 7.65E-01 | 0.00422 | ** | * |
| ICDV17: Family history of certain chronic disabling diseases | 0.42 | 2.04E-01 | 8.21E-01 | 0.01484 | * | |
| Soma: Muscle relaxant | 0.47 | 2.38E-01 | 8.75E-01 | 0.02224 | * | |
| Prednisone: Antileukemic drug | 0.54 | 3.54E-01 | 8.10E-01 | 0.00367 | ** | * |
| Iowa, Kansas, Missouri, Minnesota, Nebraska, North Dakota, South Dakota Resident | 0.54 | 3.18E-01 | 9.47E-01 | 0.02803 | * | |
| Lexapro: Selective serotonin reuptake inhibitor | 0.54 | 3.06E-01 | 9.29E-01 | 0.03124 | * | |
| ICD796: Certain adverse effects not elsewhere classified | 0.55 | 3.29E-01 | 8.67E-01 | 0.01387 | * | |
| ICD346: Migraine | 0.56 | 3.72E-01 | 8.20E-01 | 0.00394 | ** | * |
| ICD427: Cardiac dysrhythmias | 0.60 | 4.58E-01 | 7.77E-01 | 0.00014 | *** | * |
| ICD455: Hemorrhoids | 0.61 | 3.83E-01 | 9.45E-01 | 0.03179 | * | |
| ICD727: Other disorders of synovium tendon and bursa | 0.64 | 4.03E-01 | 9.76E-01 | 0.04447 | * | |
| Current Smoker | 0.68 | 5.05E-01 | 9.08E-01 | 0.00976 | ** | |
| ICD733: Other disorders of bone and cartilage | 0.69 | 5.39E-01 | 8.80E-01 | 0.00308 | ** | * |
| ICD462: Acute pharyngitis | 0.69 | 5.09E-01 | 9.36E-01 | 0.01865 | * | |
| Calcium channel blocker, Dihydropyridine Calcium Channel Blocker | 0.73 | 5.51E-01 | 9.69E-01 | 0.03118 | * | |
| Ciprofloxacin: Quinolone antibiotic | 0.74 | 5.41E-01 | 9.90E-01 | 0.04578 | * | |
| ICD530: Diseases of esophagus | 0.76 | 6.30E-01 | 9.18E-01 | 0.00448 | ** | * |
| Thiazide | 0.80 | 6.53E-01 | 9.87E-01 | 0.03863 | * | |
| ICDV70: Routine general medical examination at a health care facility | 0.81 | 6.70E-01 | 9.84E-01 | 0.03492 | * | |
| ICD477: Allergic rhinitis | 0.81 | 6.75E-01 | 9.77E-01 | 0.02844 | * | |
| Total number of medications prescribed | 0.98 | 9.68E-01 | 9.96E-01 | 0.01387 | * | |
| Total number of laboratory tests performed | 0.98 | 9.72E-01 | 9.94E-01 | 0.00266 | ** | * |
| Age | 1.02 | 1.01E+00 | 1.03E+00 | 0.00000 | *** | * |
| Total number of specialized risk factors | 1.16 | 1.08E+00 | 1.24E+00 | 0.00005 | *** | * |
| Gender: Male | 1.18 | 1.01E+00 | 1.36E+00 | 0.03160 | * | |
| Atorvastatin: Statin | 1.28 | 1.01E+00 | 1.60E+00 | 0.03744 | * | |
| Simvastatin: Statin | 1.40 | 1.12E+00 | 1.75E+00 | 0.00348 | ** | * |
| Rosuvastatin: Statin | 1.43 | 1.03E+00 | 1.97E+00 | 0.03094 | * | |
| ICD466: Acute bronchitis and bronchiolitis | 1.47 | 1.21E+00 | 1.78E+00 | 0.00009 | *** | * |
| ICD327: Organic sleep disorders | 1.48 | 1.13E+00 | 1.95E+00 | 0.00475 | ** | * |
| Carvedilol: Beta blocker, Alpha blocker | 1.54 | 1.00E+00 | 2.35E+00 | 0.04785 | * | |
| ICD257: Testicular dysfunction | 1.54 | 1.05E+00 | 2.25E+00 | 0.02710 | * | |
| Lisinopril: angiotensin-converting enzyme inhibitor activity | 1.61 | 1.33E+00 | 1.96E+00 | 0.00000 | *** | * |
| ICD428: Heart failure | 1.62 | 1.13E+00 | 2.33E+00 | 0.00879 | ** | |
| Pravastatin: Statin | 1.68 | 1.07E+00 | 2.64E+00 | 0.02428 | * | |
| ICDV04: Need for prophylactic vaccination and inoculation against certain viral diseases | 1.73 | 1.33E+00 | 2.23E+00 | 0.00003 | *** | * |
| ICD681: Cellulitis and abscess of finger and toe | 1.76 | 1.05E+00 | 2.91E+00 | 0.03045 | * | |
| Viral and chlamydial infection in conditions classified elsewhere and of unspecified site | 1.82 | 1.00E+00 | 3.16E+00 | 0.04167 | * | |
| ICD009: Ill-defined intestinal infections | 1.88 | 1.15E+00 | 3.02E+00 | 0.01036 | * | |
| ICD585: Chronic kidney disease | 1.91 | 1.45E+00 | 2.53E+00 | 0.00001 | *** | * |
| ICD053: Herpes zoster | 1.92 | 1.18E+00 | 3.11E+00 | 0.00835 | ** | |
| Hyperlipidemia Diagnosis | 1.93 | 1.67E+00 | 2.24E+00 | 0.00000 | *** | * |
| Lovastatin: Statin | 1.96 | 1.21E+00 | 3.14E+00 | 0.00558 | ** | |
| ICDV85: Body mass index | 2.01 | 1.17E+00 | 3.39E+00 | 0.00988 | ** | |
| Enalapril: angiotensin-converting enzyme inhibitor activity | 2.02 | 1.20E+00 | 3.40E+00 | 0.00778 | ** | |
| Pregabalin: Gamma-aminobutyric acid analogs | 2.09 | 1.12E+00 | 3.85E+00 | 0.01826 | * | |
| Tricor: Peroxisome Proliferator Receptor alpha Agonist | 2.16 | 1.35E+00 | 3.46E+00 | 0.00125 | ** | * |
| Venlafaxine: Serotonin–norepinephrine reuptake inhibitor | 2.22 | 1.08E+00 | 4.39E+00 | 0.02538 | * | |
| Sulfamethoxazole/trimethoprim: Miscellaneous antibiotics Sulfonamides | 2.28 | 1.36E+00 | 3.78E+00 | 0.00157 | ** | * |
| ICD302: Sexual and gender identity disorders | 2.30 | 1.37E+00 | 3.83E+00 | 0.00154 | ** | * |
| Hypertension Diagnosis | 2.46 | 2.13E+00 | 2.85E+00 | 0.00000 | *** | * |
| Cephalexin: Cephalosporin antibiotic | 2.46 | 1.25E+00 | 4.76E+00 | 0.00805 | ** | |
| Losartan: Angiotensin 2 Receptor Blocker | 2.72 | 1.69E+00 | 4.41E+00 | 0.00004 | *** | * |
| High-Risk Service Area | 6.96 | 4.72E+00 | 1.07E+01 | 0.00000 | *** | * |

In the conventional model the variables age, hypertensive status, and the interaction of age with hypertension, all were significant predictors of type 2 diabetes. The interaction between age and hypertensive status indicated that younger hypertensive patients had a greater chance of type 2 diabetes than elderly hypertensive patients. The 298 coefficients for both the full EHR model, the EHR DX Model, and the conventional model are provided within the Supplementary Data Table 1, Table 2, and Table 3.

Within the hold-out data (using the random forests probabilistic models), the EHR models had stronger sensitivity than the Conventional model for all thresholds, as shown in Table 4 and Figure 2. The AUC for the full EHR model, the EHR DX model, and the Conventional model were (81.3%, 79.6%, 74.8%). The overall accuracies were dependent on thresholds chosen, but for all thresholds, the ROC curve for the EHR models exceeded the sensitivity of the conventional model as shown in Supplementary Figure 2. Additionally, the random forest accuracies on the hold-out data were actually greater than the within-sample accuracy using the logistic regression model, demonstrating that machine-learning models have much stronger ability to predict chronic disease than the traditional parametric models. The overall accuracy of the algorithm is somewhat less important than the sensitivity and the positive predictive value. Algorithms which sought only to maximize total accuracy (removing weights and thresholds) had higher overall accuracy (84%) but markedly inferior sensitivity, implying that they would rarely classify an individual as having diabetes.

**Table 4: Models containing both diagnostic and medication information performed better than the Conventional models. While the overall accuracies depended on the choice of thresholding, for all thresholds the sensitivity of the EHR models exceeded that of conventional models.**

| Features | Model | Accuracy | Sensitivity | Specificity | Positive Predictive Value | Negative Predictive Value |
|---|---|---|---|---|---|---|
| Full EHR | Logistic Regression | 0.751 | 0.806 | 0.739 | 0.406 | 0.945 |
| DX EHR | Logistic Regression | 0.732 | 0.792 | 0.719 | 0.384 | 0.940 |
| Conventional | Logistic Regression | 0.636 | 0.780 | 0.604 | 0.304 | 0.925 |
| Full EHR | Random Forest | 0.783 | 0.624 | 0.818 | 0.431 | 0.908 |
| DX EHR | Random Forest | 0.771 | 0.603 | 0.809 | 0.411 | 0.902 |
| Conventional | Random Forest | 0.787 | 0.280 | 0.899 | 0.382 | 0.849 |

## Discussion

The EHR phenotype models outperformed the basic screening model for detecting individuals at-risk for diabetes. While many of the risk factors identified in the full EHR model have been identified previously, they have not been evaluated within the context of all other clinical factors. Our model is not prognostic; the factors we have found associated with type 2 diabetes might be complications or comorbidities of the disease, rather than its cause. However, given the finding that type 2 diabetes was diagnosed most frequently after a patient exhibited at least one complication, and that 25% of patients are undiagnosed[35], this situation may reflect clinical practice. We specifically excluded known complications of diabetes to reduce this bias, but the complications and causes of diabetes often can be intertwined (*e.g.* cardiovascular disease). Additionally, we demonstrated that the EHR had strong predictive power when excluding all physician-prescribed medications, suggesting that the diagnoses contain nearly the same information as the medications prescribed to treat them.

Each of the discussed factors is associated with diabetes and could be used as clinical markers of increased risk, but should not be considered cause or consequence of the disease. For example, specific clinic locations were associated with an increased likelihood of current type 2 diabetes. This reflects different baseline prevalence of type 2 diabetes within different communities[36]. Due to the structure of our multivariate logistic regression model, each discussed factor was significant, controlling for and separate from all other measured factors, while adjusting for the false discovery rate.

We discuss first the factors that increased the likelihood of type 2 diabetes mellitus. We do not interpret the relative magnitude of each factor, only its sign. Hypercholesterolemia and hypertension are the two other pillars of metabolic syndrome[37]; therefore we fully expected that these disorders would be highly associated due to similar underlying pathophysiology and risk factors. Medications used to treat nerve disorder have side effects of weight gain (pregabalin & gabapentin), which might indirectly be causal in the development of insulin tolerance in type 2 diabetes[38, 39]. However, because these medications also may be prescribed to treat diabetic neuropathy, this relationship was unclear. Similarly, the importance of common complications of type 2 diabetes reflects the current state of care, where patients should be assessed for type 2 diabetes if they experience complications associated with the disease, such as decreased innate or acute immune function or neural pain. Overall, this list of positively associated factors for current type 2 diabetes was in consensus with the literature. Therefore, our results suggest that EHR research can determine effectively that each of these established risk factors were independent, and meaningful in the full context of the patient.

The factors that decrease the likelihood of diabetes have been discussed less frequently in the literature, potentially because the cost of missing high-risk patients is greater than the benefit of identifying low-risk patients. Although there is no ICD-9 code for physical activity, the factors in this category reflect the treatment for mild discomfort, temporary pain, and potential athletic injuries. Athletic activity decreases insulin sensitivity, even when no weight loss is achieved, and this is important for the prevention and management of type 2 diabetes[5]. In contrast to these factors, we found also that markers of physical frailty, and diseases that inhibit food intake, decreased the risk of current type 2 diabetes. Patients with these conditions are at risk for muscle loss, which was associated negatively with type 2 diabetes[40-42]. In aggregate, these factors of increased activity or frailty can be viewed, alternatively, as signs of increased health consciousness. We found that patients who were integrated into the medical system or aware of the consequences of chronic disease were less likely to have current type 2 diabetes.

One of the benefits of EHR research is that, in addition to verifying the factors that had a relatively clear interpretation, EHR mining can identify factors where the link with disease is either unproven or unsuspected. Some factors identified here are less established in the type 2 diabetes literature. In particular, although some work has addressed the association of homosexuality, sexual identity disorders, and sexual deviancy (ICD-9 302.X) with diabetes[43, 44], our results suggest that more work should be done to understand the link between diabetes and these factors, so that at-risk patients can be identified better. In contrast to increasing the risk of current type 2 diabetes, we are uncertain why allergic rhinitis (ICD-9 477.X), and use of depot medroxyprogesterone contraceptives, decreased the prevalence of type 2 diabetes. Even though type 2 diabetes has been shown to affect innate or acute immunity[45], we are unaware of a strong link between type 2 diabetes and allergies, which primarily are a dysfunction of the adaptive or acquired immune system. Due to the underlying etiology behind the previous controversial link between exogenous hormone treatment for menopause and cardiovascular risk, we expect that the usage of depot medroxyprogesterone contraceptives reflects a patient population that engages in other activities that decrease type 2 diabetes risk, and not that depot medroxyprogesterone itself is protective for type 2 diabetes. The interpretation of each of these factors is unclear, therefore our results suggest that more work should be done to understand these observed links.

In addition to these interpretable factors, there were a number of factors that were associated negatively with type 2 diabetes, even though the literature suggested that the association should be positive. An established side effect of prednisone treatment is increased insulin resistance and steroid-induced type 2 diabetes[46]. Therefore, clinicians may prescribe prednisones only in low-risk patients. As noted above, type 2 diabetes is associated positively with disorders of the innate or acute immune system, metabolic syndrome and cardiac dysfunction; therefore we are uncertain why some of these factors were negatively associated with type 2 diabetes. Further, type 2 diabetes is a risk factor for hemorrhoids (ICD-9 455.X). Migraine also shares common comorbidities with type 2 diabetes[47], including some we identified here. Therefore, we are uncertain why the diagnosis of migraine decreased the risk of current type 2 diabetes. Our EHR model found also that being a current smoker decreased the risk of diabetes, which reflects how smoking increases base metabolic rate. Yet, this contradicts at least one study in the literature[48] which found smoking increased diabetes risk. However, this difference may be due to how we controlled for many other factors, while previous studies on smoking and diabetes accounted for a limited number of confounders (e.g. age & BMI). These factors warrant further study to understand why our analysis of EHR records did not replicate previous work.

While these results seem to summarize and review the major themes in research, we also identified factors that were not significantly associated, positively or negatively, with type 2 diabetes, and are related to the themes we discussed above. This lack of significance could reflect either a lack of statistical power due to infrequent prescribing and thereby false negatives, or a lack of an effect. Not all disorders of acute or innate immunity or the treatment for those conditions were associated with type 2 diabetes. Even though some were, not all medications for hypertension and hypothyroidism or markers of physical frailty were associated with current type 2 diabetes. Medications hypothesized to be associated with diabetes, such as varenicline[49], were not associated in our analysis.

Our model also faced challenges with missing data such as family history, and did not contain several important risk factors for diabetes, such as ethnicity and socioeconomic status[26], due to incomplete patient records. The missing data affected mainly the full EHR model, as the basic covariates (age, BMI, etc.) were rarely missing. Missing data in this population wasn't addressed, but it could have been imputed to improve accuracy and reduce bias. Patients with Type 1 diabetes were underrepresented in this sample.  Although we established a protocol for interpreting medication and laboratory test results, there was large variation in the reporting of these factors; therefore some bias and/or misreporting could be present. Due to privacy concerns, our model was unable to incorporate longitudinal information in the EHR that we would expect to improve its overall accuracy.  Moreover, this model was trained using patients who had a current diagnosis of Type 2 diabetes, which implicitly assumes that diagnosed and undiagnosed diabetes patients are similar, but this hypothesis needs to be confirmed.  In the future, we will confirm the profiles of diagnoses and undiagnosed patients using longitudinal data on an independent database, with a 12-18 month pre-index period with no diagnoses for Type I or II.

It also was intriguing that a family history of diabetes did not show a strong effect. Along with poor family medical history reporting, this implies that even although there is a genetic link for type 2 diabetes, non-genetic factors may have more effectively identified at-risk patients.  We were not able to incorporate ethnicity information into this model, because such information was unavailable. Given that the prevalence of diabetes in the sample population (18.1%) greatly exceeded the general US diabetes prevalence (11.8%), this suggests that the sample population may have had an over-representation of minority patients.  The superior performance of the EHR phenotype model compared to the conventional model may indicate then that additional health information provides additional diabetes risk factors that are especially relevant to minority populations.

**Conclusion**

Given that nearly 1 in 3 Americans will develop diabetes at some point in their lifetime, predicting and assessing diabetes risk on a population-wide scale is critical for both prevention and effective treatment. We anticipate that the largest adopter of EHR phenotype models will be insurers; risk scores will be created using existing claims databases. Given that the average patient with diabetes incurs over twice the expected cost as a patient without [7], there exists a significant financial incentive for insurers to reduce diabetes risk in patients they cover. An EHR phenotype-based pre-screen could be used for national diabetes screening at little cost, as risk scores could be computed automatically within EHRs to efficiently identify patients who should undergo more formal and sensitive laboratory screening and/or preventive behavioral interventions. Individual patients who are at high-risk for chronic and costly diseases will be targeted for patient-education programs, and reduction in calculated risk will be rewarded monetarily.

These promising results showed EHR phenotypes provided superior predictive accuracy for assessing diabetes status, compared to traditional non-laboratory information ($p<0.001$). We have demonstrated that this is possible even in the face of a diverse, at-risk patient population, with missing and incomplete patient records pooled across practice locations. This suggests that incorporating more medical history would increase the accuracy of existing diabetes risk scores in at-risk patient populations, for step-wise screening. The total increase in accuracy for full screening is partially dependent upon how much of the additional information provided by the EHR is correlated with the laboratory testing. Due to the retrospective and correlative nature of EHR research, we cannot resolve whether the EHR phenotype risk factors were a cause or effect of diabetes, but this may match clinical practice where 25% of people with type 2 diabetes are undiagnosed.

Given that this analysis showed superior prediction of diabetes using only 3 years of incomplete and unsystematically recorded data, we anticipate that the true signal and potential of an ideal, complete, and systematically utilized EHR is much greater than we demonstrate here. As EHRs become more widely used, we hope that the size and quality of the records will increase by orders of magnitude. This demands that future databases should be constructed in such a way as to allow easy data mining; and thereby to encourage researchers to develop prospective risk models[27, 50]. Reducing the amount of missing data would only strengthen the ability of these models to detect type 2 diabetes.

We, and others, have advocated that data mining EHRs could be used to address numerous clinical challenges[51-53], such as identifying patients at risk for depression, suicide, strokes, and cardiovascular events. Risk scores, not limited just to diabetes, should be automatically computed and included in a general patient profile, providing physicians an instant assessment of potential health conditions. EHR phenotypes additionally could be used for assessment of treatment efficacy, when incorporated into statistical models such as a propensity score [28].

Beyond prediction, using EHR-phenotype models provide invaluable information about the risk factors themselves. For example, EHRs can be used to assess comparative risks of medication classes (as is being done[54]), and answer important treatment questions, including whether statins or fibrates should be used to treat high cholesterol in patients with diabetes. However, our experience with hormonal therapy for menopause[55] taught us that, while there is great potential in retrospective, observational studies, the highest level of evidence is a double-blinded, randomized clinical trial. Because of this, resulting associations with diabetes are necessarily ambiguous, with no defined causal relationship, and need to be carefully scrutinized in light of complementary controlled studies[27]. For screening of a disease that is both the fifth leading cause of preventable death in the United States and has a tremendous rate of undiagnosed patients, we argue that the accuracy of the model is more important any causal inference of the predictors. Because the EHR and data mining are both nascent, there is a vast unexplored territory with applications so broad they have yet to be defined.

**Acknowledgements:**

We acknowledge NIH R33DA026109 to M.S.C., the University of California (UCLA)-California Institute of Technology Medical Scientist Training Program (NIH T32 GM08042), the Systems and Integrative Biology Training Program at UCLA (NIH T32 GM008185), and the UCLA Department of Biomathematics for partial funding of this research. We additionally thank Practice Fusion, Inc. for providing data publicly for research. Ariana E. Anderson, Ph.D., holds a Career Award at the Scientific Interface from the Burroughs Wellcome Fund.

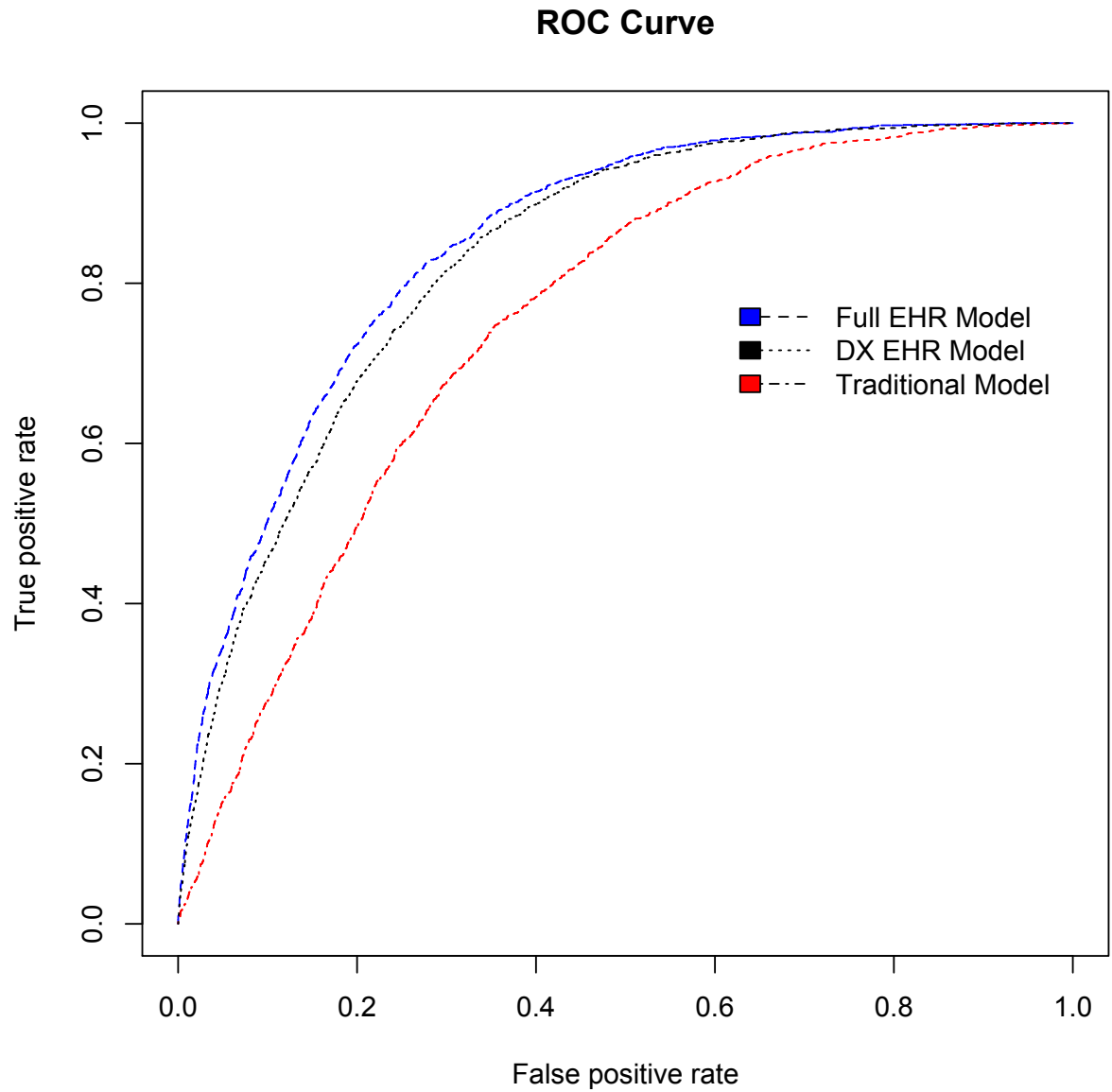

**Figure 1:** ROC Curve comparing Traditional to EHR phenotype Diabetes Prediction, using logistic regression model.

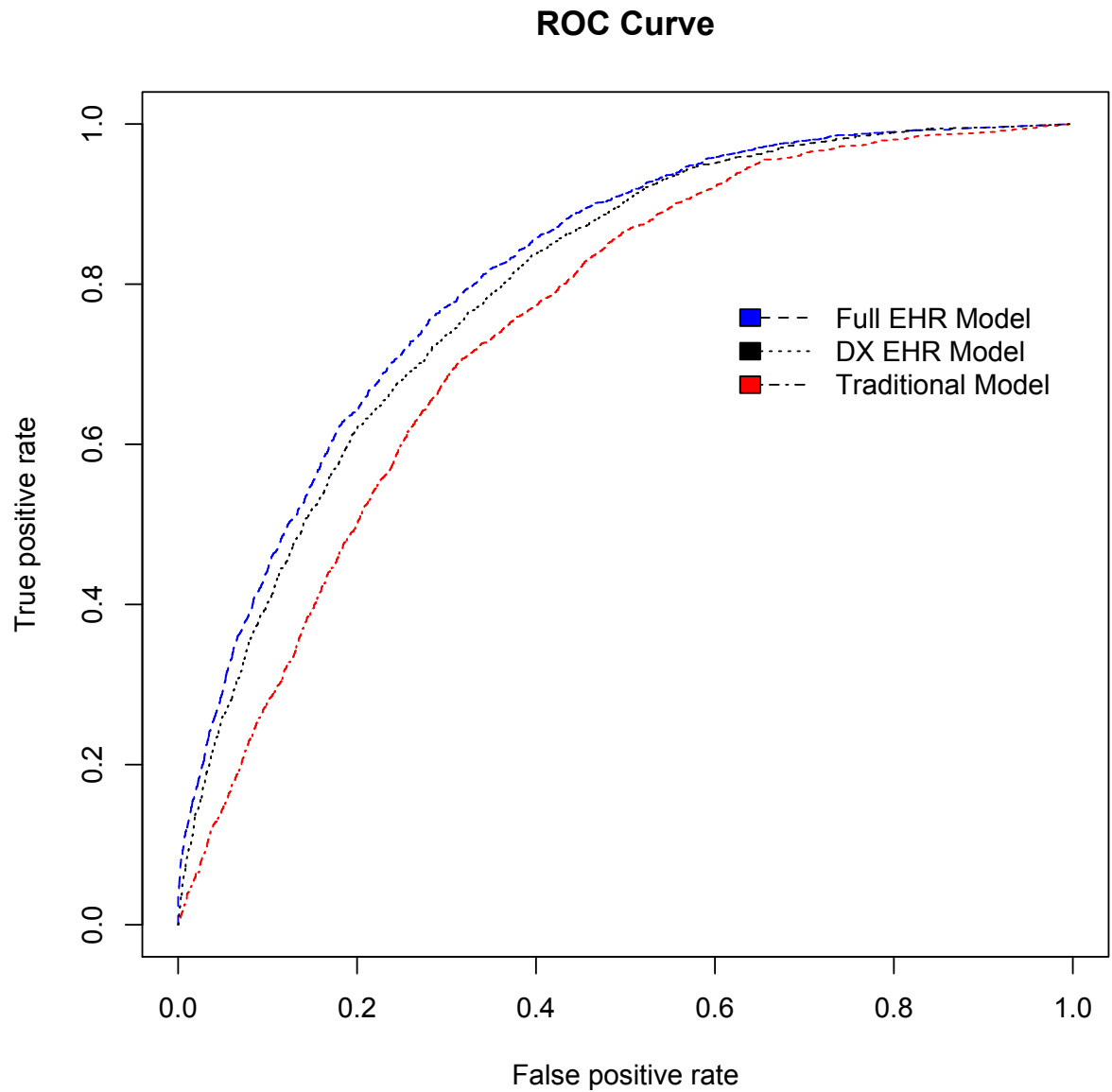

**Figure 2: ROC Curve comparing Traditional to EHR phenotype Diabetes Prediction, using random forests holdout models.**

**Competing Interests:**
The authors declare that they have no competing interests.

**Authors Contributions:**
AA conceived of the study, performed the statistical analysis, and wrote the first draft of the manuscript. WK interpreted results and models. All authors contributed to manuscript revisions, and have read and approve of the final draft.